\documentclass[a4paper]{article}

\usepackage{INTERSPEECH_v2}

\title{A Study on Replay Attack and Anti-Spoofing for Automatic Speaker Verification}
\name{Lantian Li, Yixiang Chen, Dong Wang, Thomas Fang Zheng*}
\address{
  Center for Speech and Language Technologies, Research Institute of Information Technology \\
  Department of Computer Science and Technology, Tsinghua University, Beijing, 100084, China}
\email{\{lilt13,wangdong99\}@mails.tsinghua.edu.cn, fzheng@tsinghua.edu.cn
}

\begin{document}

\maketitle
\begin{abstract}

  For practical automatic speaker verification (ASV) systems, replay attack poses a true risk. By replaying
  a pre-recorded speech signal of the genuine speaker, ASV systems tend to be easily fooled. An effective
  replay detection method is therefore highly desirable.
  In this study, we investigate a major difficulty in replay detection: the over-fitting problem
  caused by variability factors in speech signal. An F-ratio probing tool is proposed and three
  variability factors are investigated using this tool: speaker identity, speech content and playback \& recording device.
  The analysis shows that device is the most influential factor that contributes the highest over-fitting risk.
  A frequency warping approach is studied to alleviate the over-fitting problem, as verified on the
  ASV-spoof 2017 database.

\end{abstract}

\section{Introduction}

It is widely acknowledged that most biometric authentification systems are vulnerable to spoofing attacks, and anti-spoofing has become a serious focus of researchers as early as in the 1990s.
Among various biometric authentification methods, automatic speaker verification (ASV) using speech signals
is particularly interesting. Firstly, speech signals are
easy to collect, which makes ASV suitable in a very wide range of applications, especially in remote scenarios; secondly,
speech interface is easy to use and less intrusive, hence applicable to people in most circumstances;
thirdly, ASV is an active authentification (i.e., requires users' active response) thus naturally robust against spoofing.

Nevertheless, researchers indeed found some spoofing attacks which are dangerous to ASV systems,
including impersonation, speech synthesis, voice conversion, and replay~\cite{evans2013spoofing}.
In the past few years, a multitude of research on anti-spoofing has been conducted, mainly concentrated on attacks by synthetic speech
and converted speech~\cite{wu2014asvspoof, wu2015asvspoof}. Less has been done for attacks by
replayed speech, i.e., speech recorded from the genuine speaker and replayed by an imposter during the authentification.
Ironically, recording \& replay presents a more practical risk compared to speech synthesis and conversion,
as it requires neither specific expertise nor sophisticated equipments: a mobile phone is enough.
Some research has demonstrated that the low-effort replay attack may cause a much higher false-acceptance rate
than voice conversion and speech synthesis which often require complicated skills and much more effort~\cite{alegre2014re}.
In this year, the ASV-Spoof 2017 challenge~\cite{kinnunen2017asvspoof} put its focus on replay attack, which may
invoke more interest in this area.

In this study, we investigate how the replay detection is so difficult. Our hypothesis is that this difficulty
can be largely attributed to the intermingling of the distortion caused by
the recording \& replay process and some other variability factors in speech signals,
including speaker identity, speech content and playback \& recording device. To test this hypothesis, we propose to use
the F-ratio metric as a probing tool to analyze the impact of variability factors.
The analysis shows that device is the most influential factor: the F-ratio patterns
are very different among different devices, which will significantly reduce the generalizability of replay detection
models, leading to high risk of over-fitting. To alleviate this problem, a frequency warping approach is studied.
The key idea is to emphasize on
the most discriminative frequencies while deemphasizing the frequencies that are less discriminative
but impacted by the variation of devices.

The rest of this paper is organized as follows. Section~\ref{sec:rel} discusses some related work
on replay attack and countermeasures, and Section~\ref{sec:db} describes the ASV-Spoof database we used
in the  study. Section~\ref{sec:f-ratio} presents the variation analysis using the F-ratio probing tool,
and Section~\ref{sec:detect} presents the replay detection results. The paper is concluded in Section~\ref{sec:conl}.

\section{Related work}
\label{sec:rel}

An early work related to replayed speech in ASV research was conducted by Villalba and colleagues~\cite{villalba2010speaker}. They investigated
the vulnerability of ASV systems with far-field replayed speech. Using an ASV system based on joint factor analysis (JFA),
they reported an increase in the equal error rate (EER) from $1\%$ to almost $70\%$ when real-time speech of imposters
was replaced by replayed speech of genuine speakers.
The same authors later showed that it was possible to detect such replay attack by measuring the
difference of channels with and without replay~\cite{villalba2011preventing}.
Alegre et al.~\cite{alegre2014re} re-assessed the risk of replay attack to ASV.
They conducted the experiments with six ASV systems using the large and standard NIST corpus.
Their results showed that the low-effort replay attack posed a significant risk on all the
tested ASV systems, and the state-of-the-art i-vector/PLDA system was the most vulnerable.

Shocked by the real threat from replay attack, researchers have carried out a number of studies in recent two years.
Wu et al.~\cite{wu2014study} evaluated the vulnerability of text-dependent ASV systems
under replay attack using a standard benchmarking database, and also proposed
a replay detection technique to safeguard the verification system.
The key idea of the detection algorithm was to match the input signal with some previously stored
speech samples based on a similarity metric.
Another replay detection approach was proposed by Villalba and colleagues~\cite{villalba2011detecting}. They
assumed that a far-field recorded speech
signal would involve more noise and reverberation, which, as a consequence, would result in
a flattened spectrum and reduced modulation indexes within the signal.
Following this assumption, they designed several features that were argued to be suitable for detecting
far-field replayed speech.

Previous studies have confirmed that replay attack is a genuine risk to ASV systems,
and some replay detection methods have been proposed with certain success. However, our
knowledge to the properties of replayed speech remains limited, and the replay detection is still difficult.

\section{ASV-Spoof Database}
\label{sec:db}

The ASV-Spoof database, which was released for the ASV-Spoof 2017 challenge, was used in our experiments.
The full database contains three subsets: training set, development set and evaluation set.
The database is primarily based on the recent text-dependent \emph{RedDots} corpus~\cite{lee2015reddots} and its replayed version~\cite{kinnunen2017reddots}.
It contains 10 common phrases, and the recording was conducted with different playback \& recording devices.
The sampling rate is 16 kHz and the sample precision is 16 bits.
As the ground truth of evaluation set has not been published, we just use the training and development set in our experiments.
Table~\ref{tab:data} presents more details of the database.

    \begin{table*}[htp]
    \begin{center}
        \caption{Data Profile of ASV-Spoof database}
        \label{tab:data}
          \begin{tabular}{|l|c|c|c|c|c|c|}
           \hline
                 Database       &   No. of Spks   &   Playback Devices   &  Recording Devices  &  No. of Genuine Utts.  &  No. of Spoofed Utts.    \\
           \hline
           \hline
                 Training     &      10         &           3          &           1         &       1,508           &        1,508               \\
           \hline
                 Development    &      8          &           6          &           7         &        760            &        950                 \\
            \hline
          \end{tabular}
    \end{center}
    \end{table*}

\section{Variation analysis by F-ratio}
\label{sec:f-ratio}

The difficulty of replay detection is multiple, but a particular difficulty may be in
the fact that the spectral distortion caused by the recording and replay process
is intermixed with the multiple variability factors in speech signals,
such as speaker identity, speech content and playback \& recording device.
This intermixing means a replay detection model trained in one condition would be unsuitable
for other conditions, i.e., over-fitting. This is particularly the case if the training data is
limited and thus the variation of the variability factors can not be fully covered.
We will testify this hypothesis in this section.

\subsection{F-ratio probing tool}

In order to test the factor-mixing hypothesis, the F-ratio metric~\cite{wolf1972efficient} is used as a probing tool to
test the impact of the variability factors. The usage of F-ratio is because it directly relates to
the detection task, so the analysis results are more valuable for feature and model design.

F-ratio is a simple frequency-domain weighting
approach where the weight of a frequency band is determined by its discriminative capability for the task in hand.
In speaker recognition, this technique has been used in feature design, i.e., promoting the most significant
frequency bands in the front-end pipeline~\cite{wang2016improving}. In our study, the F-ratio values can be used to
investigate which frequency bands are more capable to separate genuine (real-time)
and replayed speech.

Formally, suppose the spectrum of the speech signal is divided into $M$ filter banks (Fbanks). The
F-ratio value of the $i$-th Fbank is defined as the ratio of the between-class distance and the within-class
variance of this Fbank values of all the frames, where the classes are genuine speech ($C_g$) and replayed speech ($C_r$).
This is formulated as follows:

\vspace{-2mm}
\[
    F_i = \frac{(\mu_{i}^g-\mu_{i}^r)^2}{\frac{1}{N_g}\sum_{x_i \in C_g}(x_{i}-\mu_{i}^g)^2 + \frac{1}{N_r}\sum_{x_i \in C_r}(x_{i}-\mu_{i}^r)^2}
\]

\noindent where $x_i$ represents the value of the $i$-th Fbank of the speech frame $x$, and $\mu_{i}^g$ and $\mu_{i}^r$ are
the means of $x_i$ of all the frames of the genuine speech class and the replayed speech class, respectively. $N_g$ and $N_r$ are the number of frames of the two classes.

The F-ratio values $[F_1, F_2, ..., F_M]$ form an F-ratio pattern. This pattern will be heavily used in this paper to
analyze the properties of replayed speech. The simplest usage of the F-ratio pattern is to discover which frequency
bands are most discriminative for detecting replayed speech; more importantly, the variation of the F-ratio patterns caused
by a particular variability factor will reflect the impact of the factor to replay detection in generalizability.

\subsection{Variation analysis by F-ratio probing}
\label{sec:prob}

In this section, we use the F-ratio pattern as a probing tool to investigate the
impact of variability factors in the replay detection task. Three factors are
considered: speaker identity, speech content and playback \& recording device.
The experiment is conducted on the development set of the ASV-Spoof database
because it contains more playback \& recording devices than the training set.

\begin{figure*}[htb]
    \centering
    \includegraphics[width=0.85\linewidth]{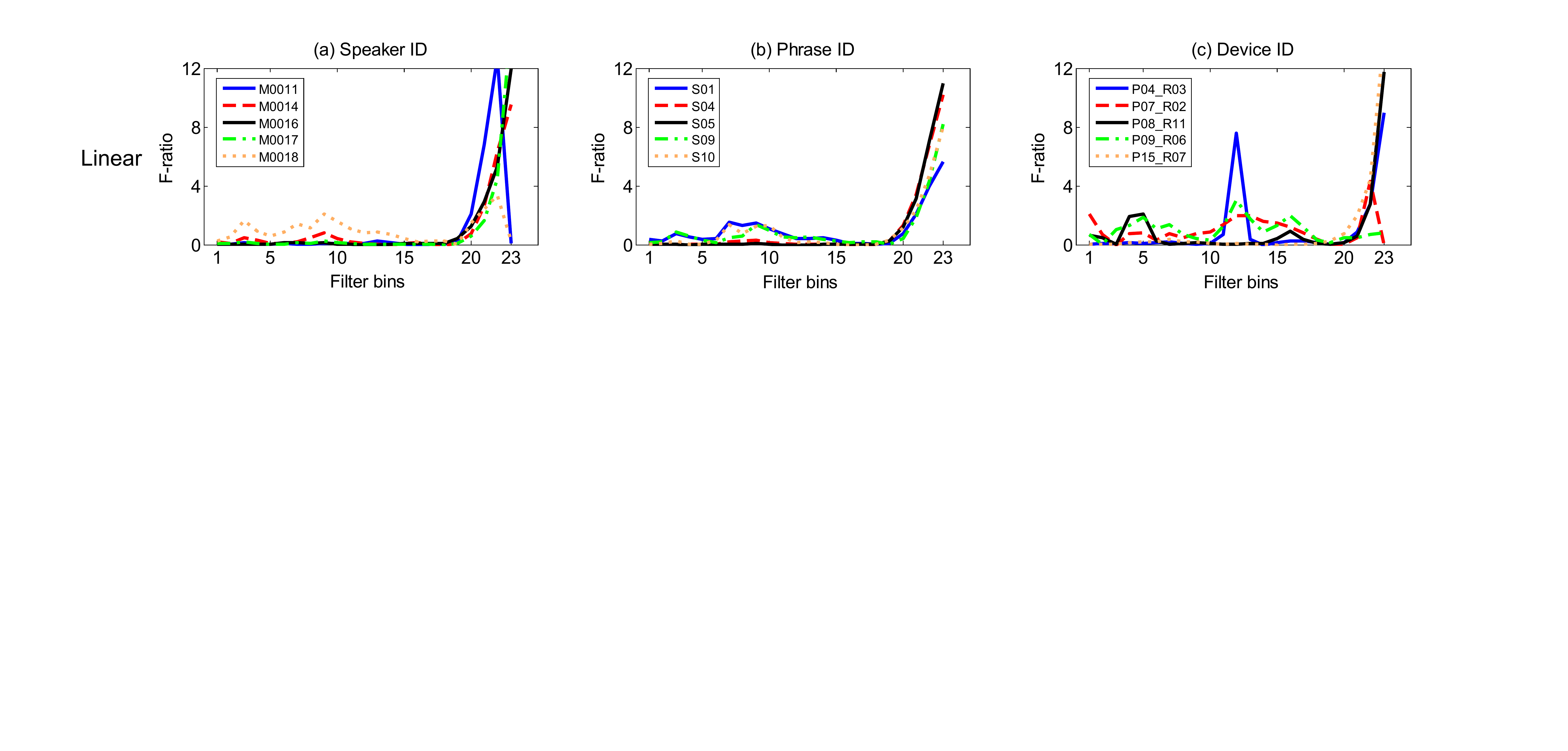}
    \vspace{-1mm}
    \caption{F-ratio patterns on linear Fbanks.}
    \vspace{-2mm}
    \label{fig:fratio-A}
    \end{figure*}

In each of the experiments, we select one of the three factors as the probed
factor. By choosing every value of the probed factor and drawing the
F-ratio patterns for all of them,
we can check the variation of the F-ratio patterns caused by this factor.
The results are shown in Figure~\ref{fig:fratio-A}, where the three plots
use (a) speaker identify (Speaker ID), (b) speech content (Phrase ID), (c) playback \& recording device (Device ID) as
the probed factor, respectively. In each experiment, the frequency axis is
linearly divided into $23$ Fbanks (with equal bandwidth),
and the pattern for each value of the probed factor is presented as a curve
in the corresponding plot of Figure~\ref{fig:fratio-A}.

Figure~\ref{fig:fratio-A} (a) and (b) reveal two important messages: Firstly,
the most discriminative information resides in the high-frequency bands;
secondly, the F-ratio patterns are rather similar in spite of the values of
the speaker identity and the speech content. If we regard the F-ratio pattern as a simple model,
this similarity means that a model trained in one condition of speaker or speech content
is suitable for data in other conditions. In other words, the variations of
speaker and speech content do not impact model generalizability.

Figure~\ref{fig:fratio-A} (c) delivers more interesting information. From this plot, we
observe very different F-ratio patterns with different playback \& recording devices. This
indicates that with individual devices, the discovered F-ratio patterns are rather
device-dependent, or, in other words, not generalizable.
This also implies serious over-fitting in model training (with features derived from the Fbanks),
since discriminative information discovery and usage is general to all classification models.
Note that the weak generalizability of replay detection models
has been noticed and warned by the organizers of the ASV-Spoof 2017 challenge.

\subsection{Compensation by frequency warping}

The analysis of the pervious section suggests that a major difficulty
in replayed detection is the weak generalizability of features/models
caused by the variation of devices. A pre-processing
that can reduce this variation will help improve model generalizability
and the performance of replay detection.

A key observation from Figure~\ref{fig:fratio-A} is that in these three plots,
the high-frequency bands show great discriminative capability. Particularly
in Figure~\ref{fig:fratio-A} (c), the discrepancy of F-ratio patterns with different
devices is mostly in the low- and middle- frequency bands. This means that if we can
emphasize the high-frequency bands, then the device-related variation may be
largely reduced, while most discriminative power still remains.

\begin{figure}[htb]
    \centering
    \includegraphics[width=0.9\linewidth]{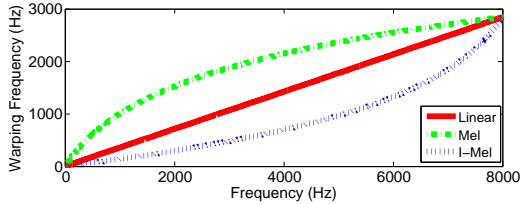}
    \vspace{-1mm}
    \caption{The forms of three warping functions.}
    \vspace{-3mm}
    \label{fig:feature}
\end{figure}

We use the frequency warping approach to achieve the goal.
In this approach, the frequency axis is firstly warped by a nonlinear function,
and then the Fbanks are designed on the warped frequencies. A famous
warping approach is the Mel warping, which emphasizes on low-frequency bands by
`stretching' the frequency axis in the low-frequency area and `compressing' it in the
high-frequency area. This leads to the popular Mel Fbanks (M-Fbanks).
Since our goal is to emphasize the high-frequency bands, we study an
inverted Mel warping function~\cite{chakroborty2009improved} to achieve the effect, which leads to a new
design of Fbanks denoted by IM-Fbanks. For a clear presentation, the original (linear) Fbanks without any
frequency warping is denoted by L-Fbanks.
%The equations of three warping functions are shown in Equ.~\ref{eq:ln}
%, and the function forms are illustrated in Figure~\ref{fig:feature}.
The forms of three warping functions are illustrated in Figure~\ref{fig:feature}.

%\vspace{-3mm}

%\begin{equation}
%\label{eq:ln}
%\left\{
%       \begin{array}{rcl}
%             $Linear$ (f) & = & 0.35 * f \\
%             $Mel$ (f) & = & 1127 * log (1 + f / 700) \\
%             $I-Mel$ (f) & =  & 2840 - 1127 * log (1 + 8000 - f ) / 700. \\
%       \end{array}
%\right.
%\end{equation}

\vspace{-4mm}

%\vspace{-2mm}

\subsection{Probe M-Fbanks and IM-Fbanks}

\begin{figure*}[t]
    \centering
    \includegraphics[width=0.85\linewidth]{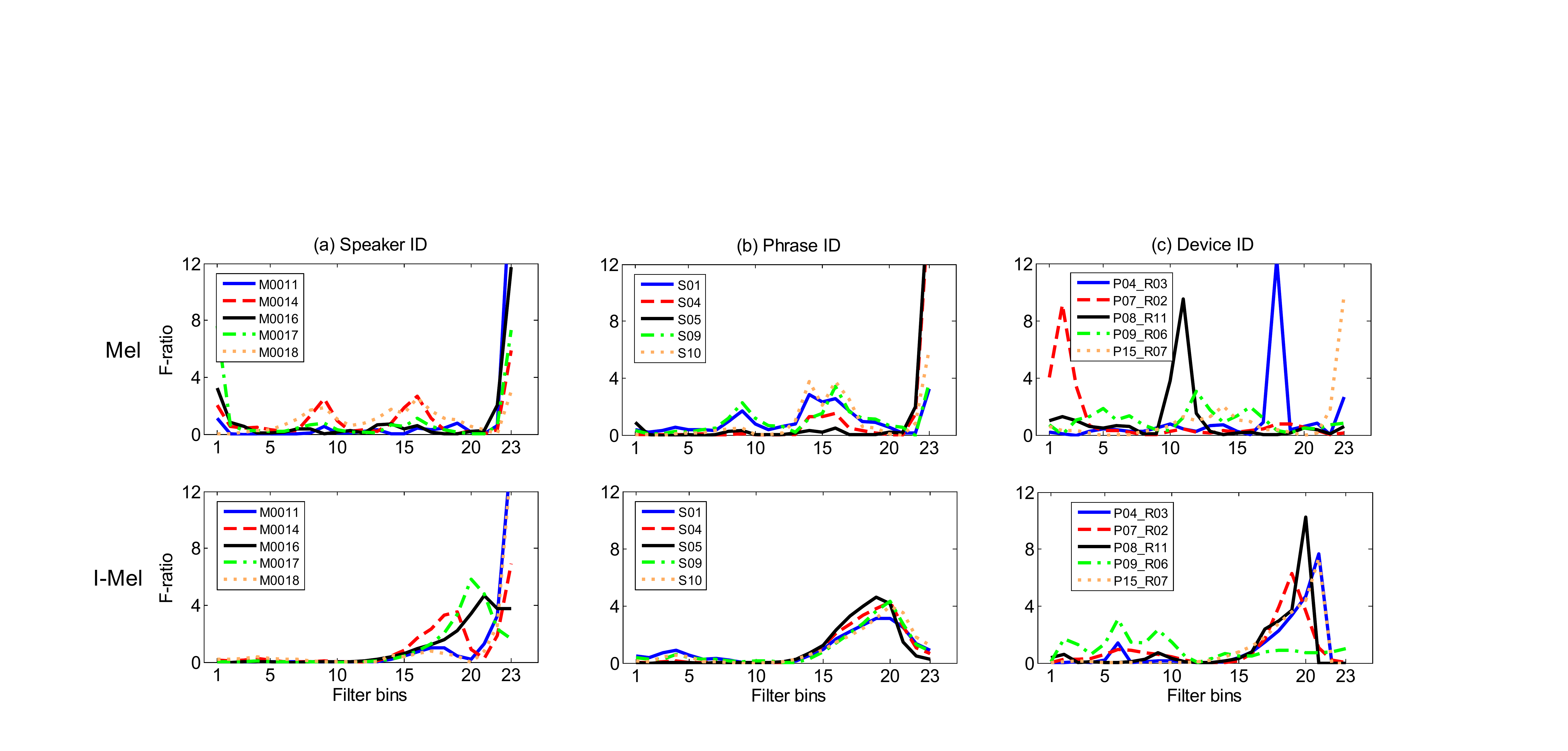}
    \vspace{-1mm}
    \caption{F-ratio patterns on M-Fbanks and IM-Fbanks.}
    \vspace{-2mm}
    \label{fig:fratio-B}
    \end{figure*}

We apply the F-ratio probing tool to test the variations caused by the three
variability factors with M-Fbanks and IM-Fbanks, following the same process in Section~\ref{sec:prob}.
The results are shown in Figure~\ref{fig:fratio-B}, where the first row presents the results
with M-Fbanks, and the second row presents the results with IM-Fbanks.

It can be seen that with the M-Fbanks (emphasizing on low frequencies), the device-related
variation becomes more significant, while with the IM-Fbanks (emphasizing on
high frequencies), the variation is largely reduced. This variation reduction is highly
important for building models that can be well generalized during test. These results
double confirmed our analysis that the most discriminative information for replay detection
is within high-frequency bands and the variation of devices mostly impacts the low-frequency
bands. By emphasizing on high-frequency bands (as in IM-Fbanks), the variation can be reduced
while keeping the discriminative power.

\vspace{-1mm}

\subsection{Probe training and development sets}

\begin{figure}[htb]
    \centering
    \includegraphics[width=0.9\linewidth]{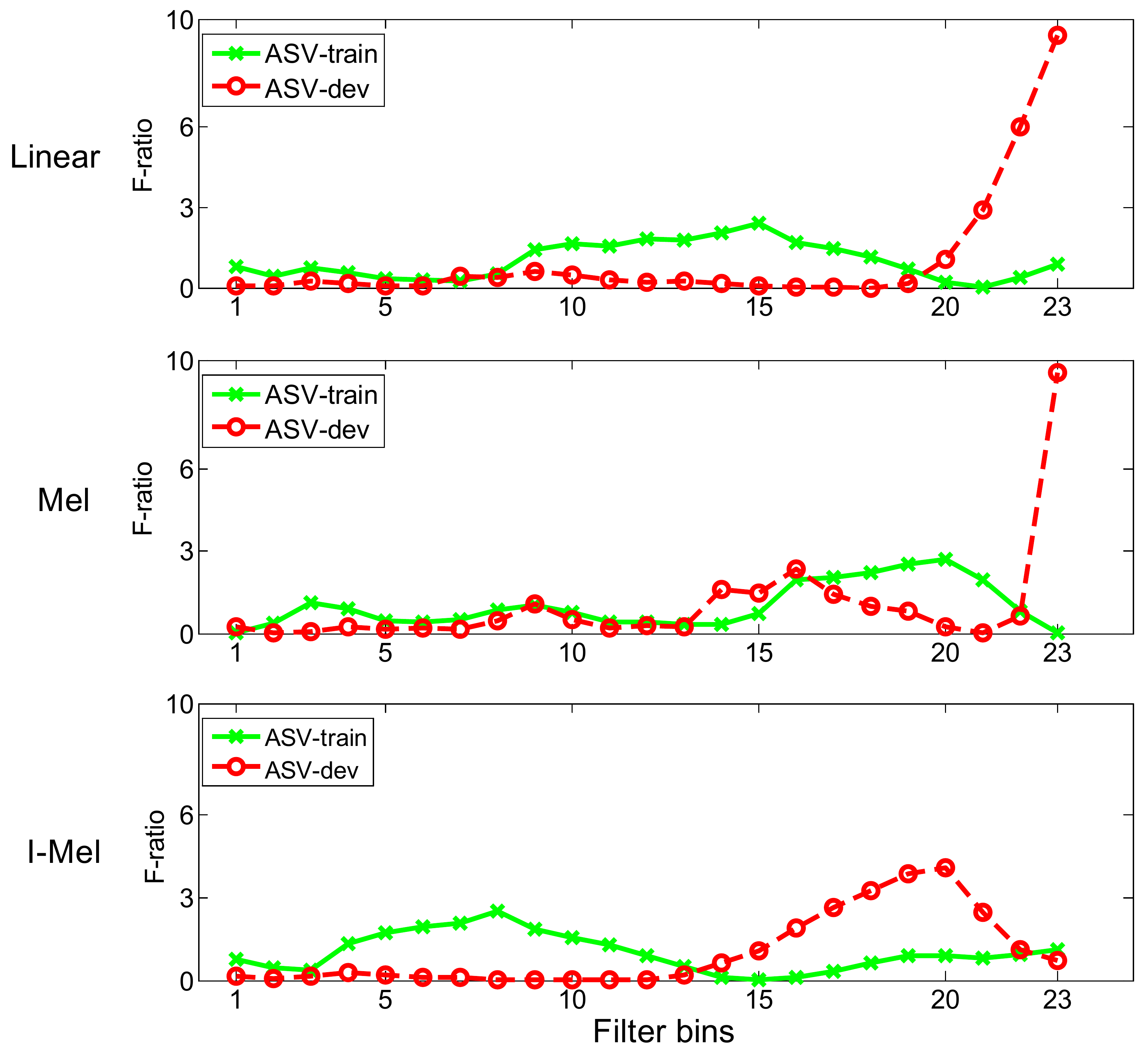}
    \vspace{-1mm}
    \caption{F-ratio patterns on the training and development sets.}
    \vspace{-3mm}
    \label{fig:fratio-C}
\end{figure}

In the last experiment of this section, we study the discrepancy between the training and development sets using the
F-ratio probing tool. The results are shown in Figure~\ref{fig:fratio-C}, where the F-ratio patterns are
computed from both the training and development sets, using each of the three types of Fbanks. It can be observed that the
F-ratio patterns on the two sets are quite different, indicating that models trained on the training
data may be not generalizable well on the development data.
Interestingly, the F-ratio patterns of the two sets with the IM-Fbanks are much closer
when compared to those with the other two types of Fbanks. This means that the IM-Fbanks
can improve model generalizability.

\section{Replay detection}
\label{sec:detect}

This section utilizes these three Fbanks to conduct replay detection. We firstly present the
classifiers used in the experiment, and then report the detection results.

\subsection{Classifiers}

We experiment with three classifiers: GMM, i-vector/SVM, and DNN. GMM
is a pure generative model and DNN is a pure discriminative model. I-vector/SVM
is a combination of the two types of models. In general, generative models are more
generalizable than discriminative models, while discriminative models are more powerful
if the class boundary is complicated.

For the DNN model, we use the 23-dimensional Fbanks as the input feature.
While for the GMM and i-vector/SVM, a discrete Fourier transform (DCT) is applied to form
a 13-dimensional cepstral feature, and then plus its first order derivatives, resulting in 26-dimensional feature vectors.
For the M-Fbanks, this results in the standard Mel-frequency cepstral coefficients (MFCC).
Following the name convention, the cepstral feature derived from the L-Fbanks is denoted by LFCC, and
the one derived from the IM-Fbanks is denoted to be IMFCC.

\subsubsection{GMM system}

For the GMM system, we first train two GMMs, one for genuine speech (denoted by $\lambda_{g}$) and one for replayed
speech (denoted by $\lambda_{r}$).
During test, a sequence of feature vectors ${X}$ are presented to the
two GMMs, and the log likelihood ratio $\Lambda(X)$ is used to make the genuine/replay
decision, where the likelihood ratio is defined as follows:

\vspace{-4mm}
\[
  \Lambda(X) = \frac{1}{N}\sum_{i=1}^N {\left\{log p(x_{i}|\lambda_{g}) - log p(x_{i}|\lambda_{r})\right\}}.
\]

\subsubsection{i-vector/SVM system}

For the i-vector/SVM system, an i-vector model is firstly trained using the training data, including
both genuine and replayed speech. This model is then used to extract
the representations of all the training utterances, and this representation is a low-dimensional
continuous vector which is called \emph{i-vector}.
Once the i-vectors of all the training utterances are ready, an SVM model is trained for
genuine/replay classification.

During test, the input utterance is embedded into an i-vector using the i-vector model,
and then the SVM model is applied to determine whether the utterance is genuine or replayed.

\subsubsection{DNN system}

The DNN model used in this experiment consists of a two-layers convolutional neural network (CNN) for frame-level
feature extraction and a time-delay neural network (TDNN) for frame-level classification.
During test, given an input utterance, the posteriors for the genuine and replay classes are computed frame by frame,
and then these posteriors are averaged to produce an utterance-level posterior, which is then used for
genuine/replay decision making.

\subsection{Replay detection results}
\label{sec:system}

We built three replay detection systems as described above using the training data,
with three features (MFCC, LFCC, IMFCC). The detection results in terms of equal error rate (EER) on
the development set are shown in Table~\ref{tab:exp}.

It can be observed that on the ASV-Spoof database, the IMFCC is the most effective. This is
expected as IMFCC leads to less over-fitting, according to the results of the F-ratio variation analysis.
When comparing different models, the best
performance is obtained with the simple GMM model. For DNN, although the results seem acceptable, but
we found the model is not reliable: different initialization will result in rather different results, and
some are quite poor. This can be explained by the potential over-fitting
we discussed during the variation analysis, and the fact that complex models are more
prone to over-fitting.

      \begin{table}[htp]
      \begin{center}
        \vspace{-1mm}
        \caption{EER(\%) results on replay detection with different features and classifiers. `GMM(Diag)' means
        GMM with diagonal covariance; `GMM(Full)' means GMM with full covariance; `SVM(Rbf)' means SVM with RBF kernel;
        `SVM(Linear)' means SVM with linear kernel. }
        \vspace{-1mm}
        \label{tab:exp}
          \begin{tabular}{|c|c|c|c|}
            \hline
            System                  &  MFCC      &    LFCC       &   IMFCC     \\
            \hline
            GMM(Diag)               &  24.87     &   13.42       &   8.85       \\
            GMM(Full)               &  23.55     &   12.37       &   7.50   \\
            \hline
            i-vector + SVM(Rbf)     &  28.82     &   17.37       &   9.34      \\
            i-vector + SVM(Linear)  &  26.32     &   17.37       &   10.92    \\
            \hline
            CT-DNN                  &  16.45     &   8.16        &   8.29    \\
            \hline
          \end{tabular}
      \end{center}
      \end{table}
      \vspace{-6mm}

\section{Conclusions}
\label{sec:conl}

In this paper, we used the F-ratio metric as a probing tool to analyze the impact of
various speech factors to replay detection. Our analysis showed
the most discriminative information for replay detection resides in high-frequency bands;
moreover, the playback \& recording device contributes the most variation in F-ratio patterns,
hence the most significant over-fitting risk. A frequency warping approach
was studied to alleviate the variation caused by devices, and significant
performance gains were obtained. Future work involves better frequency compensation
methods and analysis for between-dataset discrepancy.

\section{Acknowledgements}

This work was supported by the National Natural Science Foundation of China under Grant No. 61371136 / 61633013
and the National Basic Research Program (973 Program) of China under Grant No. 2013CB329302.

\bibliographystyle{IEEEtran}

\bibliography{mybib}

\begin{thebibliography}{10}
\providecommand{\url}[1]{#1}
\csname url@samestyle\endcsname
\providecommand{\newblock}{\relax}
\providecommand{\bibinfo}[2]{#2}
\providecommand{\BIBentrySTDinterwordspacing}{\spaceskip=0pt\relax}
\providecommand{\BIBentryALTinterwordstretchfactor}{4}
\providecommand{\BIBentryALTinterwordspacing}{\spaceskip=\fontdimen2\font plus
\BIBentryALTinterwordstretchfactor\fontdimen3\font minus
  \fontdimen4\font\relax}
\providecommand{\BIBforeignlanguage}[2]{{%
\expandafter\ifx\csname l@#1\endcsname\relax
\typeout{** WARNING: IEEEtran.bst: No hyphenation pattern has been}%
\typeout{** loaded for the language `#1'. Using the pattern for}%
\typeout{** the default language instead.}%
\else
\language=\csname l@#1\endcsname
\fi
#2}}
\providecommand{\BIBdecl}{\relax}
\BIBdecl

\bibitem{evans2013spoofing}
N.~W. Evans, T.~Kinnunen, and J.~Yamagishi, ``Spoofing and countermeasures for
  automatic speaker verification.'' in \emph{INTERSPEECH}, 2013, pp. 925--929.

\bibitem{wu2014asvspoof}
Z.~Wu, T.~Kinnunen, N.~Evans, and J.~Yamagishi, ``Asvspoof 2015: Automatic
  speaker verification spoofing and countermeasures challenge evaluation
  plan,'' 2014.

\bibitem{wu2015asvspoof}
Z.~Wu, T.~Kinnunen, N.~Evans, J.~Yamagishi, C.~Hanil{\c{c}}i, M.~Sahidullah,
  and A.~Sizov, ``Asvspoof 2015: the first automatic speaker verification
  spoofing and countermeasures challenge,'' \emph{Training}, vol.~10, no.~15,
  p. 3750, 2015.

\bibitem{alegre2014re}
F.~Alegre, A.~Janicki, and N.~Evans, ``Re-assessing the threat of replay
  spoofing attacks against automatic speaker verification,'' in
  \emph{Biometrics Special Interest Group (BIOSIG), 2014 International
  Conference of the}.\hskip 1em plus 0.5em minus 0.4em\relax IEEE, 2014, pp.
  1--6.

\bibitem{kinnunen2017asvspoof}
T.~Kinnunen, N.~Evans, J.~Yamagishi, K.~A. Lee, M.~Sahidullah, M.~Todisco, and
  H.~Delgado, ``Asvspoof 2017: automatic speaker verification spoofing and
  countermeasures challenge evaluation plan,'' 2017.

\bibitem{villalba2010speaker}
J.~Villalba and E.~Lleida, ``Speaker verification performance degradation
  against spoofing and tampering attacks,'' in \emph{FALA workshop}, 2010, pp.
  131--134.

\bibitem{villalba2011preventing}
------, ``Preventing replay attacks on speaker verification systems,'' in
  \emph{Security Technology (ICCST), 2011 IEEE International Carnahan
  Conference on}.\hskip 1em plus 0.5em minus 0.4em\relax IEEE, 2011, pp. 1--8.

\bibitem{wu2014study}
Z.~Wu, S.~Gao, E.~S. Cling, and H.~Li, ``A study on replay attack and
  anti-spoofing for text-dependent speaker verification,'' in
  \emph{Asia-Pacific Signal and Information Processing Association, 2014 Annual
  Summit and Conference (APSIPA)}.\hskip 1em plus 0.5em minus 0.4em\relax IEEE,
  2014, pp. 1--5.

\bibitem{villalba2011detecting}
J.~Villalba and E.~Lleida, ``Detecting replay attacks from far-field recordings
  on speaker verification systems,'' in \emph{European Workshop on Biometrics
  and Identity Management}.\hskip 1em plus 0.5em minus 0.4em\relax Springer,
  2011, pp. 274--285.

\bibitem{lee2015reddots}
K.-A. Lee, A.~Larcher, G.~Wang, P.~Kenny, N.~Br{\"u}mmer, D.~A. van Leeuwen,
  H.~Aronowitz, M.~Kockmann, C.~Vaquero, B.~Ma \emph{et~al.}, ``The reddots
  data collection for speaker recognition.'' in \emph{INTERSPEECH}, 2015, pp.
  2996--3000.

\bibitem{kinnunen2017reddots}
T.~Kinnunen, M.~Sahidullah, M.~Falcone, L.~Costantini, R.~G. Hautamaki,
  D.~A.~L. Thomsen, A.~K. Sarkar, Z.-H. Tan, H.~Delgado, M.~Todisco
  \emph{et~al.}, ``Reddots replayed: A new replay spoofing attack corpus for
  text-dependent speaker verification research,'' in \emph{IEEE International
  Conference on Acoustics, Speech and Signal Processing (ICASSP)}.\hskip 1em
  plus 0.5em minus 0.4em\relax IEEE, 2017.

\bibitem{wolf1972efficient}
J.~J. Wolf, ``Efficient acoustic parameters for speaker recognition,''
  \emph{The Journal of the Acoustical Society of America}, vol.~51, no.~6B, pp.
  2044--2056, 1972.

\bibitem{wang2016improving}
L.~Wang, J.~Wang, L.~Li, T.~F. Zheng, and F.~K. Soong, ``Improving speaker
  verification performance against long-term speaker variability,''
  \emph{Speech Communication}, vol.~79, pp. 14--29, 2016.

\bibitem{chakroborty2009improved}
S.~Chakroborty and G.~Saha, ``Improved text-independent speaker identification
  using fused mfcc \& imfcc feature sets based on gaussian filter,''
  \emph{International Journal of Signal Processing}, vol.~5, no.~1, pp. 11--19,
  2009.

\end{thebibliography}

\end{document}